\documentclass[
    twocolumn, 
]{aastex7}

\usepackage{microtype}
\usepackage{amsmath}

\providecommand{\sorthelp}[1]{}

\begin{document}

\title{The End of the Road for Far-infrared Reddening Maps?\\ Evidence for Reddening Errors Driven by Changes in PAH Abundance}

\author[0000-0002-3455-1826, gname=Dennis, sname=Lee]{Dennis Lee}
\affiliation{Jet Propulsion Laboratory, California Institute of Technology, 4800 Oak Grove Drive, Pasadena, CA 91109, USA}
\altaffiliation{\textcopyright~2025 California Institute of Technology. Government sponsorship acknowledged.}
\email[hide]{dennisl@jpl.nasa.gov}

\author[0000-0001-7449-4638, gname=Brandon, sname=Hensley]{Brandon S. Hensley}
\affiliation{Jet Propulsion Laboratory, California Institute of Technology, 4800 Oak Grove Drive, Pasadena, CA 91109, USA}
\email[hide]{brandon.s.hensley@jpl.nasa.gov}

\author[0000-0001-5929-4187]{Tzu-Ching Chang}
\affiliation{Jet Propulsion Laboratory, California Institute of Technology, 4800 Oak Grove Drive, Pasadena, CA 91109, USA}
\affiliation{California Institute of Technology, 1200 E. California Boulevard, Pasadena, CA 91125, USA}
\email[hide]{tzu-ching.chang@jpl.nasa.gov}

\author[0000-0001-7432-2932, gname=Olivier, sname=Dore]{Olivier Dor\'e}
\affiliation{Jet Propulsion Laboratory, California Institute of Technology, 4800 Oak Grove Drive, Pasadena, CA 91109, USA}
\affiliation{California Institute of Technology, 1200 E. California Boulevard, Pasadena, CA 91125, USA}
\email[hide]{olivier.p.dore@jpl.nasa.gov}


\begin{abstract}
Accurate correction for extinction by Galactic dust is essential for studying the extragalactic sky.
In the low-extinction regions of the Ursa Major molecular cloud complex, we demonstrate that Galactic dust reddening maps constructed from observations of far-infrared emission are insensitive to variations in the abundance of polycyclic aromatic hydrocarbons (PAHs), and, as a result, to PAH-induced variations in reddening.
Using galaxy counts to validate various reddening maps, we find evidence that maps based on far-infrared emission erroneously under-predict reddening compared to stellar reddening maps.
This underestimation by far-infrared emission based reddening maps---representing the largest discrepancy between maps of up to $E(B-V)=0.08$ mag---is correlated with the relative brightness of PAH emission.  
Furthermore, we demonstrate theoretically that changes in PAH abundance via accretion from the gas phase is capable of altering extinction significantly with only minor changes to far-infrared emission.
We show that modeling the extinction of Ursa Major using both far-infrared and mid-infrared emission more accurately traces dust extinction variations due to changes in PAH abundance.
Finally, we discuss how SPHEREx observations of the 3.3\,$\mu$m PAH feature are a promising way to overcome this limitation of far-infrared emission.
\end{abstract}

\section{Introduction}\label{sec:intro}

Studies of the extragalactic Universe are invariably affected by the presence of dust in the Milky Way.
Dust grains scatter and absorb electromagnetic radiation at optical and near-infrared wavelengths and re-emit in the infrared as thermal radiation.
Inferring the intrinsic brightness and color of any astrophysical source requires the application of a dust extinction correction to the observed brightness and colors.

Dust extinction can be traced using a range of different methods and a variety of different datasets. 
As a result, a wide array of dust reddening maps are available.
While each of these techniques seeks to determine the same quantity, each has its own set of advantages and limitations.

The thermal emission of dust at far-infrared wavelengths provides one avenue to estimate the dust reddening.
Such maps are built on the assumption that the extinction can be inferred from the total dust column and that the total dust column can be derived from the far-infrared emission, typically employing multiple bands to account for dust temperature effects.
In general, extinction maps derived from far-infrared emission benefit from the higher angular resolution and all-sky coverage of the maps. 
Observations from satellites such as the Infrared Astronomical Satellite \citep[IRAS;][]{Neugebauer:1984}, the Cosmic Background Explorer \citep[COBE;][]{Boggess:1992}, and Planck \citep{planck2016-l01} have enabled the production of dust maps with all-sky coverage \citep[e.g.,][]{Schlegel:1998, planck2013-p06b, planck2016-XLVIII}.
The most widely used reddening map---\citet[][]{Schlegel:1998} (``SFD'')---is produced using emission at far-infrared wavelengths as measured by IRAS and DIRBE. 
However, reddening maps based on far-infrared dust emission are known to suffer from contamination from the cosmic infrared background \citep[e.g.,][]{Yahata:2007, Chiang:2019b}.

Measurements of optical reddening toward background stars provide a direct measurement of extinction \citep[e.g.,][]{Schlafly:2011, Green:2014, Green:2015, Green:2019, Edenhofer:2024, Zhou:2024, Zhang:2025a}.
However, in regions of low extinction (such as those targeted by cosmological surveys), direct stellar reddening measurements can become dominated by factors such as uncertainties in the star's intrinsic spectra or color as these become comparable to the magnitude of the dust reddening.
Overall, these techniques are limited by the location and distance of stars at any given sight line.
Sight lines with fewer visible stars, whether intrinsic or due to too much dust obscuration, suffer from reduced statistical sampling.
Since this method can only measure the dust extinction out to the distance of the star, depending on the distance of the available stars, it may or may not be possible to measure the full Galactic dust extinction.
This can be particularly limiting for extragalactic and cosmological studies requiring the correction of all the dust from the Milky Way. 

Finally, as dust is typically well mixed with the neutral gas in the interstellar medium \citep[see, e.g.,][]{Boulanger:1996}, dust can also be traced using measurements of the \ion{H}{1} 21-cm line. 
This was employed in the construction of early dust maps by \citet{Burstein:1978, Burstein:1982} with a more recent application by \citet{Lenz:2017}.
These methods are limited in usefulness in high-density regions (typically at lower Galactic latitudes) where the gas fraction of the interstellar medium with more molecular or ionized hydrogen is higher, though this can be mitigated if the total hydrogen column density is known \citep{Bohlin:1978, Nguyen:2018}. Even at higher Galactic latitudes there are significant but poorly understood discrepancies between \ion{H}{1}- and far-infrared based extinction maps \citep{Cheng:2025}.

Current and upcoming cosmological surveys aiming to precisely constrain cosmological parameters demand accurate correction for Galactic extinction.
Recently, \citet{Karim:2025} conducted a cross-correlation analysis of the Dark Energy Spectroscopic Instrument (DESI) emission line galaxy sample with CMB lensing to measure the amplitude of the matter power spectrum $\sigma_8$.
This cosmological parameter is of particular interest due to the statistically significant disagreement that has been observed when inferred by different experiments \citep[e.g.,][]{Abbott:2020, Pandey:2022}.
When the reddening map used to model the foreground systematics was varied, \citet{Karim:2025} discovered that this had a significant impact on their final estimated value of $\sigma_8$ and thus to the nature of tensions with other experiments (e.g., Planck).
As such, our cosmological constraints are limited by the systematics of various reddening maps, requiring an improved understanding of any uncertainties and biases involved in their construction.

Current models of dust extinction at optical wavelengths invoke contributions from both small (radius of $\sim10^{-3}$\,$\mu$m) and large dust grains (radius of $\sim10^{-1}$\,$\mu$m) \citep[e.g.,][]{Hensley:2023, Ysard:2024}. 
Submicron grains are comparable in size to the wavelengths of interest, and so are especially efficient at scattering.
These grains likely dominate the optical extinction.

However, small grains also contribute to the optical extinction via the prominent 2175\,\AA\ extinction ``bump.'' 
As little scattering is observed in this feature \citep{Lillie:1976} and as an absorption feature in this wavelength range is found in hydrocarbon materials like graphite \citep{Stecher:1965, Draine:1989}, it is widely attributed to carbonaceous nanoparticles, such as polycylcic aromatic hydrocarbons \citep[PAHs;][]{Joblin:1992}.
The feature is quite broad \citep[FWHM $\sim$500\,\AA;][]{Fitzpatrick:1986}, and so its strength has a non-negligible impact on the optical extinction.

Accurate estimation of the dust extinction therefore requires accounting for both grain populations, which may vary with respect to each other across the sky.
Indeed, the relative abundance of PAHs and larger submicron grains has been observed to spatially vary both in other galaxies \citep[e.g.,][]{Sandstrom:2010, Aniano:2012, Aniano:2020, Chastenet:2019, Chastenet:2023, Spilker:2023, Whitcomb:2024, Chastenet:2025} as well as in our own \citep[e.g.,][]{Mattila:1996, Kaneda:2014, planck2013-p06b, Hensley:2016}.
In addition to these larger scale, modest variations, more dramatic changes in PAH abundance have been observed in localized regions, such as \ion{H}{2} regions and molecular clouds \citep[e.g.,][]{Sandstrom:2010, Egorov:2023}.
A recent study of PAHs has attributed variations observed in the slope of the optical extinction curve near several Milky Way molecular clouds to growth of PAHs via accretion \citep{Zhang:2025a}.

In this work, we investigate a potential limitation of commonly used far-infrared based extinction maps that originates in far-infrared emission's inability to trace the abundance of PAHs.
Changes in PAH abundance can result in changes to both the total extinction and its wavelength dependence.
Since PAHs do not emit strongly in the far-infrared, such variations will not be captured by reddening maps based solely on far-infrared emission.
On the other hand, PAHs do emit strongly in the mid-infrared.
As a result, incorrect extinction estimates from employing only far-infrared emission may be ameliorated by incorporating measurements of mid-infrared emission that better trace PAH abundance.

We conduct our investigation toward the Ursa Major cloud located in the North Celestial Pole Loop \citep{Heiles:1984, Heiles:1989, Meyerdierks:1991b}.
This region is among the clouds noted for its PAH abundance variations in \citet{Zhang:2025a} and is also the region with the largest discrepancy between the DESI and SFD reddening maps \citep{Zhou:2024}.
Despite being located high above the Galactic Plane, it is also a region of relatively high column density \citep{Marchal:2023}.
Ultimately, we leverage the discrepancies observed in this region to identify limitations of current extinction maps and explore avenues for improving future ones.

This paper is organized as follows:
Section~\ref{sec:data} presents the various reddening maps we investigate. 
Section~\ref{sec:galaxy_backlight_test} details the emission line galaxy backlight test we use to validate the performance of the different reddening maps.
Section~\ref{sec:correlation} describes the performance of the various reddening maps in the context of PAH abundance.
Section~\ref{sec:astrodust_pah_model} demonstrates theoretically how dust emission and extinction evolve with increasing PAH abundance.
Section~\ref{sec:rv_modeling} details our modeling of the extinction in Ursa Major and the contributions from the mid-infrared and the far-infrared.
Section~\ref{sec:discussion} discusses the implications of our analysis as well as promising avenues of addressing the limitations of far-infrared emission with SPHEREx.
Finally, we conclude in Section~\ref{sec:conclusion}.

\section{Data}\label{sec:data}
In our study of the extinction in the Ursa Major cloud, we employ reddening maps that broadly fall into two categories: maps derived directly from photometry or spectroscopy of stars and maps derived indirectly via mid-/far-infrared dust emission.

\subsection{Stellar Reddening Maps}
\label{subsec:stellar_reddening_maps}

Reddening maps can be constructed by comparing the observed photometry or spectroscopy of stars against a prediction of their intrinsic spectra, typically from stellar atmosphere models or through comparison with unreddened stars of similar spectral type. The reddening due to the intervening dust along a given sight line can then be directly determined, and reddenings to many stars assembled into maps, including in the radial dimension since the distances to many stars are known. 
Here, we consider two recent stellar reddening maps constructed using reddening measurements toward large samples of stars:

\begin{enumerate}
    \item \citet[][DESI]{Zhou:2024} produced Galactic reddening maps by directly measuring the extinction toward 2.6 million stars. 
    Stellar spectroscopy from the first two years of the DESI spectroscopic survey was used to model the stellar spectra and produce synthetic colors (with no extinction).
    Reddening in the $g-r$ and $r-z$ bands was then computed by comparing these synthetic colors with the observed colors of stars from DESI imaging.
    For all analysis in the present work, we use the $E(B-V)$ reddening map derived from the more reliable $g-r$ colors (as opposed to from $r-z$).
    We use the reddening measurements binned into a HEALPix\footnote{\href{http://healpix.sourceforge.net}{http://healpix.sourceforge.net}}~\citep{Gorski:2005}~map of $N_{\rm side}=128$ (angular resolution of $\approx27'.5$) for the highest completeness in the Ursa Major cloud.
    
    \item \citet[][Gaia]{Zhang:2025b} used Gaia XP spectra obtained from Gaia's two low-resolution prism spectrographs BP and RP to measure the dust toward $\sim$130 million stars.
    Combining this with near-infrared photometry, \citet{Zhang:2025b} forward modeled the XP spectra using a neural network model trained on the subset of the Gaia XP data with accompanying high-resolution LAMOST spectroscopy \citep{Cui:2012, Zhao:2012}. 
    This work modeled both the extinction as well as the extinction curve resulting in measurements of both reddening $E(B-V)$ and $R_V$.
    $R_V$ is the single parameter commonly used to describe the extinction curve at optical to near-infrared wavelengths.
    We construct the 2D Gaia $E(B-V)$ and $R_V$ maps using the binning method .
    To ensure completeness in the Ursa Major region, we choose a HEALPix pixelization of $N_{\rm side}=256$ (angular resolution of $\approx13'.7$).
    We select stars beyond 1.0 kpc to capture the full Galactic extinction along the line of sight and apply the same cuts as described in \citet{Zhang:2025b}.
    These general quality cuts exclude stars with poor fitting and unreliable effective temperature $T_{\rm eff}$ estimates.
    The $T_{\rm eff}$ cut, in particular, excludes unreliable determinations of the slope in the inferred stellar spectra and the resulting unreliable estimates of $R_V$.
\end{enumerate}

\subsection{Indirect Infrared Emission-Based Reddening Maps}
\label{subsec:emission_reddening_maps}

\subsubsection{Far-Infrared Emission}

Dust grains in the interstellar medium are heated by starlight and radiate thermally in the far-infrared. 
As a result, this diffuse far-infrared emission traces the interstellar dust column density, enabling the construction of dust reddening maps.
However, dust grains can be heated to a range of different temperatures depending on the intensity of the local radiation field, resulting in varying emissivities. 
To account for this variation, conversion from infrared emission to column density---and thus to reddening---requires information about the dust temperature. 
Here, we consider three maps relying on far-infrared emission ($\lambda \geq 100$\,$\mu$m ):

\begin{enumerate}
    \item The most widely used reddening map has been \citet[][SFD]{Schlegel:1998}, primarily constructed using the all-sky 100\,$\mu$m~map from IRAS.
    To correct for the temperature-dependent emissivity, \citet[][]{Schlegel:1998} used the 100\,$\mu$m~and~240\,$\mu$m~maps from the Diffuse Infrared Background Experiment (DIRBE) instrument on COBE to produce a dust temperature map ($T_{\rm SFD}$).
    This temperature map was then used to apply a dust temperature correction to the IRAS 100\,$\mu$m~emission map to produce a temperature-corrected 100\,$\mu$m~emission map with an angular resolution of $6'.1$. 
    To obtain the reddening, the temperature-corrected IRAS 100\,$\mu$m~emission was then calibrated to the observed reddening of external elliptical galaxies. 

    \item Using Planck and DIRBE/IRAS observations, \citet[][MF15]{Meisner:2015} modeled the dust emission from 100~GHz (3000\,$\mu$m) to 3000~GHz (100\,$\mu$m) as the sum of two modified blackbodies.
    Such a model can represent two physically different dust populations (e.g., silicate and carbonaceous grains).
    To produce a reddening map, \citet{Meisner:2015} used the stellar sample from \citet{Schlafly:2011} to empirically calibrate the dust optical depth from their model fit to the observed reddening\footnote{\texttt{lambda\_meisner\_finkbeiner\_2015\_dust\_map.fits}}.
    
    \item We implement a simple reddening map based on the Planck 857~GHz (350\,$\mu$m) emission from the Planck Public Release 3 \citep[][Planck-857]{planck2016-l03}\footnote{\texttt{HFI\_SkyMap\_857\_2048\_R3.01\_full.fits}}. 
    To account for the dust emissivity, we apply a temperature correction using the SFD dust temperature map $T_{\rm SFD}$ \citep{Schlegel:1998, Finkbeiner_Schlegel_Davis_2016}.
    We then convert the temperature-corrected Planck 857~GHz emission in MJy\,sr$^{-1}$ to $E(B-V)$ reddening values by deriving a linear conversion factor to SFD of $E(B-V)/I_{\nu, {857}} = 1.95\times10^{-2}$ mag (MJy\,sr$^{-1}$)$^{-1}$. 
    The resulting reddening map has a spatial resolution of approximately $5'$. 
\end{enumerate}

\subsubsection{Mid-Infrared Emission}
Numerous emission features originating from PAHs have been observed in the infrared from approximately 3\,$\mu$m~to 20\,$\mu$m~\citep{Leger:1984}.
While PAHs represent only the smaller dust grains in the interstellar medium, their emission is nonetheless a potentially useful indirect tracer of dust reddening.

\citet[][WISE-W3]{Meisner:2014} processed the diffuse emission observed in the Wide-field Infrared Survey Explorer \citep[WISE;][]{Wright:2010} Band~3, centered at roughly 12~$\mu$m, into a full sky map in the mid-infrared\footnote{\texttt{wssa\_sample\_1024-bintable.fits}}. 
The WISE~W3 band spans numerous PAH features, including the 8.6, 11.2, and 12.7\,$\mu$m~features as well as parts of the 7.7 and 16.4\,$\mu$m~features.
After removal of compact sources, imaging artifacts, and other contamination, the final map is calibrated to a zero point based on the Planck 857~GHz emission and smoothed to a resolution of $15"$. For comparison with the other reddening maps, we use the downsampled $3'.4$ resolution $N_{\rm side}=1024$ HEALPix map. \

To convert the emission in MJy~sr$^{-1}$ to reddening values, we begin by applying a temperature-correction to the measured emission to account for varying emissivities.
Unlike far-infrared emission where the emission is proportional to the Planck function evaluated at a steady-state temperature, mid-infrared emission arises from smaller dust grains that are stochastically heated by ambient photons.
As a result, the mid-infrared emission is proportional to the energy density of the interstellar radiation field (ISRF). 
Since the steady-state temperature $T$ of dust grains that dominate the far-infrared emission is proportional to the ISRF intensity $u$ by:
\begin{equation}
    u \propto T^{4 + \beta},
\end{equation}
with dust opacity spectral index $\beta$, we can compute the temperature-corrected WISE~W3 emission using:
\begin{equation}
    I_{\nu}^{\rm W3, T} =  I_{\nu}^{\rm W3} \times \left(\frac{T}{T_0}\right)^{-(4 + \beta)}\,.
\end{equation}
We adopt $\beta=1.5$ \citep{planck2016-l11A} and use the SFD dust temperature map $T=T_{\rm SFD}$. 
We use $T_0=17.9$\,K, the median of $T_{\rm SFD}$ for the whole sky, though we note that the resulting reddening map created with this temperature correction is independent of the exact chosen value for $T_0$.
Finally, we derive a linear conversion factor from the \citet{Meisner:2014} emission in MJy~sr$^{-1}$ to $E(B-V)$ to the SFD reddening map.
We find $E(B-V)/I_{\nu, {\rm W3}} = 5.63\times10^{-1}$ mag (MJy~sr$^{-1}$)$^{-1}$.

\begin{figure*}
    \centering
    \includegraphics[width=1\textwidth]{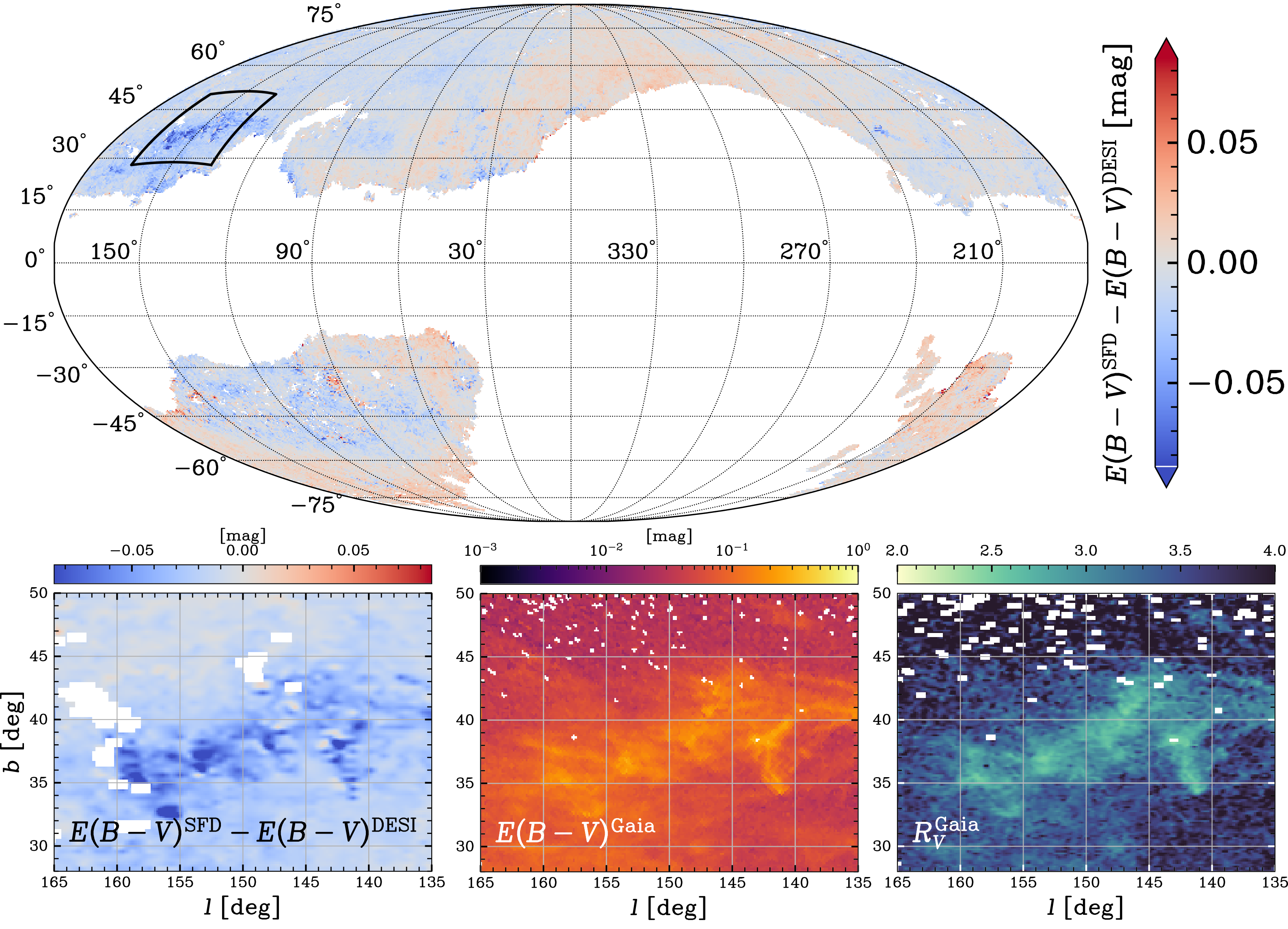}
    \caption{\textit{Top:} Difference between the SFD and DESI reddening maps in the shared footprint. The black outline indicates the Ursa Major region where SFD underestimates the extinction compared to DESI. \textit{Bottom:} A zoom in of the difference the SFD and DESI reddening maps is shown on the left. The $E(B-V)$ (center) and $R_V$ (right) as measured by Gaia for the Ursa Major region. The SFD reddening map underestimates the extinction in the Ursa Major region where the value of $R_V$ is lower. \label{fig:ursa_ma_zoom}}
\end{figure*}

\subsection{\texorpdfstring{\ion{H}{1} 21-cm Emission}{HI 21-cm Emission}}
It is also possible to trace reddening by exploiting the strong correlation between \ion{H}{1} 21-cm emission and dust emission. 
However, the \ion{H}{1} 21-cm emission based reddening map from \citet{Lenz:2017} is limited to sky regions with $N_{\text{\ion{H}{1}}}< 4\times10^{21}$~cm$^{-2}$ where the relationship between $N_{\text{\ion{H}{1}}}$ and $E(B-V)$ is linear.
At higher \ion{H}{1} column densities, where the column density of molecular gas in non-negligible, this relationship becomes nonlinear.
As the Ursa Major region is largely above this threshold ($>25\%$ of pixels have $N_{\text{\ion{H}{1}}} > 4\times10^{21}$~cm$^{-2}$,  accounting for essentially all of the denser structure in the region), we do not include an \ion{H}{1} 21-cm based reddening map for comparison in this work. 

\subsection{Comparison in Ursa Major}
Figure~\ref{fig:ursa_ma_zoom} shows the difference in measured extinction between the SFD and DESI reddening maps for the entire footprint with shared measurements between SFD and DESI. 
An enlarged view of this difference in the Ursa Major region, along with the Gaia reddening $E(B-V)$ and Gaia $R_V$ in the region, are also presented.
The Ursa Major region is where the extinctions measured via DESI and SFD most strongly diverge.
Furthermore, the magnitude of the difference between maps is correlated with $E(B-V)$ as well as with $R_V$.

\section{Galaxy Backlight Test}
\label{sec:galaxy_backlight_test}

While differences exist between various extinction maps---such as those seen in Figure~\ref{fig:ursa_ma_zoom} between SFD and DESI---it is not clear which map most accurately traces the true reddening.
As the distribution of galaxies in the Universe should be isotropic, we can use galaxy counts to validate reddening maps.
Given a set of criteria for selecting a sample of extragalactic objects (such as galaxies) and a chosen Galactic extinction map to correct the photometry for Galactic dust, we can use the spatial uniformity of the resulting sample to gauge the accuracy of the chosen extinction map.
Galactic dust is expected to be the dominant systematic in photometric imaging of extragalactic objects.
As a result, the density of the galaxy sample resulting from a selection is expected to be largely spatially uniform if the chosen extinction map is accurate and the dust extinction is accurately accounted for (though we do not expect complete uniformity due to the intrinsic galaxy-density fluctuations of our universe). 

\citet{Zhou:2024} demonstrated this method by comparing the DESI and SFD reddening maps using the DESI emission line galaxy sample---specifically a modified ELG\_LOP sample \citep{Raichoor:2023}.
The term {ELG\_LOP} refers to the sample's ``low'' fiber assignment priority sample (as opposed to the separate ELG\_VLO sample with ``very low'' fiber assignment priority).
The galaxy sample is selected from the Legacy Surveys (LS) DR9 imaging catalog's $g$, $r$, and $z$-band photometry \citep{Dey:2019} and is designed to target emission line galaxies in the redshift range of $0.6 < z < 1.6$ with a focus on the $1.1 < z < 1.6$ range.
Using this galaxy sample, \citet{Zhou:2024} found that the DESI reddening map produces a galaxy density map free from systematic large-scale spatial features that were observed when using the SFD reddening map instead.
Regions of agreement between DESI and SFD generally show more uniform galaxy density while large-scale spatial features generally correspond to regions of greater disagreement between the two extinction maps~\cite[][Figure~9 and Figure~13]{Zhou:2024}.

\begin{figure*}[!hb]
\centering
\includegraphics[width=0.99\textwidth]{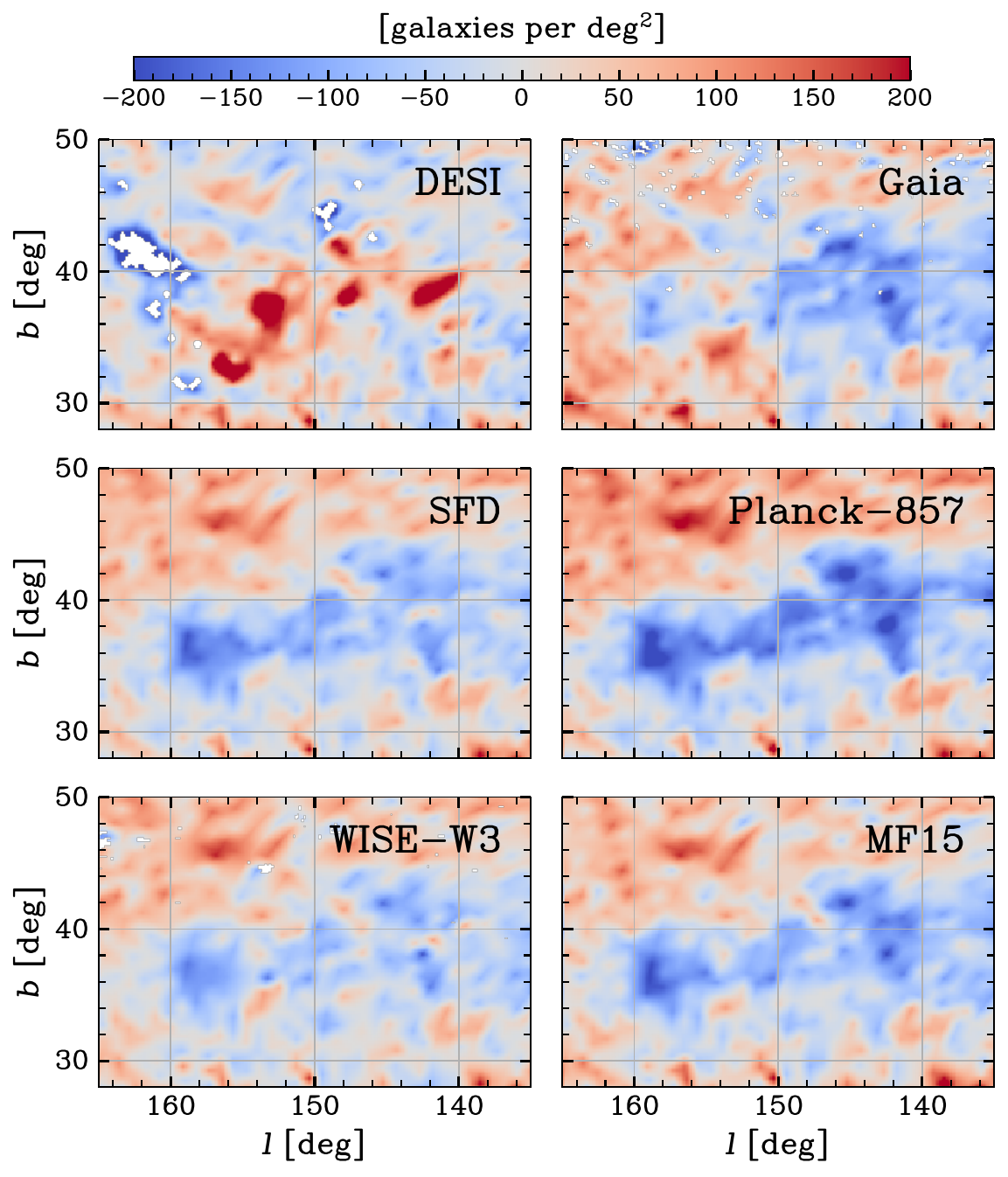}
\caption{Galaxy density per deg$^2$ of the DESI ELG\_LOP galaxy sample selected using various reddening maps. The galaxy density is median-subtracted to emphasize the degree of uniformity. The Ursa Major cloud can be seen clearly in the underestimates of the galaxy sample maps obtained using the SFD, Planck-857, and MF15 maps. On the other hand, the galaxy samples created using the DESI and Gaia reddening maps generally show less spatial correlation with the Ursa Major cloud. 
The WISE-W3, while still showing some spatial correlation, shows comparatively less when compared to the SFD, Planck-857, and MF15 maps. 
\label{fig:galaxy_selection_map}}
\end{figure*}

We extend this test to the additional reddening maps detailed in Section~\ref{sec:data}. 
Due to the range of resolutions of the various reddening maps, we resample all of the maps to a common $0'.85$ grid of a $N_{\rm side}=4096$ HEALPix map.
To obtain the galaxy sample using each reddening map, we then follow the modified {ELG\_LOP} selection criteria used in \citet{Zhou:2024}. 
Figure~\ref{fig:galaxy_selection_map} shows the ELG\_LOP galaxy sample resulting from selection by each reddening map.
We find the median ELG\_LOP galaxy density to be between 542 to 588 galaxies per deg$^2$ depending on the exact reddening map used.

Overall, each map shows a similar trend where the denser regions of Ursa Major result in an under density of galaxies. This is most evident in the galaxy density maps using the SFD, Planck-857, and MF15 reddening maps which have an under density of the ELG\_LOP galaxy sample clearly associated with the Ursa Major cloud. Such an under density is not clearly visible in the maps obtained using the DESI reddening map and substantially less visible in the galaxy sample obtained using the Gaia and WISE-W3 reddening maps. Unlike the other galaxy samples, the galaxy sample resulting from using the DESI map shows an over density of galaxies in the denser regions of Ursa Major.

As only the reddening map differs in the galaxy sample-selection procedure, we can attribute the differences in the resulting galaxy density to the different estimates of reddening in this region.
Furthermore, given that we expect uniformity in galaxy density and the residuals observed are clearly associated with the Ursa Major cloud, there is evidence that the SFD, Planck-857, and MF15 reddening maps are underestimating the true reddening in this region.
Similarly, it would appear that DESI, Gaia, WISE-W3, and MF15  maps provide estimates of dust extinction that are---when compared to SFD, Planck-857---less contaminated by Galactic structure. 
While the under-density clearly associated with the Ursa Major cloud seen in the far-infrared emission based reddening maps is the focus here, this does not imply that the other reddening maps (i.e., DESI, Gaia, WISE-W3, MF15) are free from any systematics and inaccuracies.
In fact, for example, the difference in behavior of the galaxy density maps between the DESI and Gaia reddening maps---both constructed via direct measurements of stellar extinction---reveal that each map likely has systematics beyond what is discussed here.

We can quantitatively assess the uniformity of the galaxy samples using the median absolute deviation (MAD) value.
We find that the DESI reddening map shows the lowest amount of dispersion in the galaxy sample with a median absolute deviation (MAD) value of 31 galaxies per deg$^2$.
The WISE-W3 map is also among the most uniform with a MAD value of 33 galaxies per deg$^2$.
By comparison, the SFD, MF15, and Planck-857 all show higher MAD values of 39, 39, and 55 galaxies per deg$^2$, respectively, indicating more variation.
By this metric, the Gaia extinction map with a MAD value of 44 exhibits more variations than all but Planck-857.
However, given that these variations in the galaxy density map visually do not appear as correlated with Galactic structure when compared to SFD, Planck-857, and MF15, it is more likely that whatever deficiencies that exist in the Gaia map are unrelated to the actual Galactic structure.
While the MAD values indicate the presence of some inaccuracy in the extinction maps, it alone does not inform us about the origin of systematic effects.

\begin{figure}
    \centering
    \includegraphics[width=0.99\columnwidth]{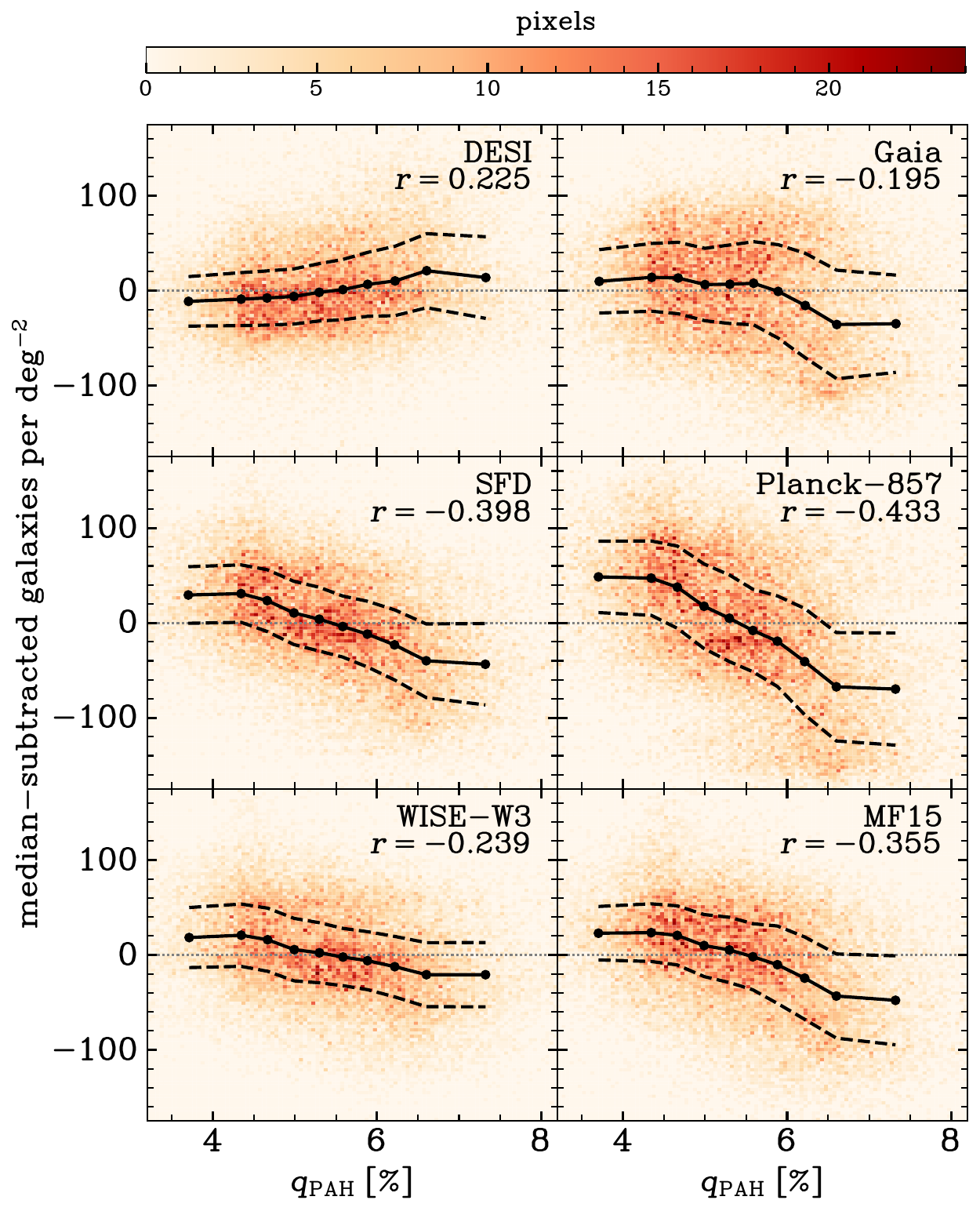}
    \caption{Median-subtracted ELG\_LOP galaxy sample density vs.\,$q_{\rm PAH}$ for each reddening map. The Pearson correlation coefficient $r$ is also shown. Both the SFD and Planck-857 reddening maps produce galaxy samples that are strongly correlated with $q_{\rm PAH}$.}
    \label{fig:galdense_v_qPAH}
\end{figure}

\section{Galaxy Sample Density and PAH Mass Fraction}
\label{sec:correlation}
We find that the deficits observed in the galaxy density maps for certain reddening maps to be correlated with the PAH mass fraction (i.e., the fraction of the total dust mass in the form of PAHs) in Ursa Major. 
To estimate the PAH mass fraction in Ursa Major, we use the PAH mass fraction $q_{\rm PAH}$ from \citet{planck2014-XXIX}. 
These $q_{\rm PAH}$ values are obtained through fitting observations from Planck, IRAS, and WISE to the dust model of \citet{Draine:2007}.

As the $q_{\rm PAH}$ values were derived in part from the \citet{Meisner:2014} all-sky W3 maps, we use the accompanying data quality flags to remove any bad or contaminated pixels.
This includes pixels contaminated from solar system objects or the Moon, saturated pixels, linelike effects, and latent ghosts that originate from the instrument electronics as well as regions that suffer from low integer frame coverage (corresponding to bit flags 0, 3, 8, 15, 18, 20, and 21).

Figure~\ref{fig:galdense_v_qPAH} plots the galaxy density as a function of $q_{\rm PAH}$ for the various reddening maps.
The SFD, Planck-857, and MF15 reddening maps all exhibit a clear negative correlation (Pearson $r = -0.398, -0.433, -0.355$, respectively) where areas of higher PAH mass fraction coincide with a deficit in galaxy sample density. 
On the other hand, direct measurements of reddening show weaker correlations with $q_{\rm PAH}$, with the DESI reddening map showing a weak positive correlation ($r=0.225$) and the Gaia and WISE-W3 maps showing weak negative correlations ($r=-0.195$ and $-0.239$, respectively).

In regions of Ursa Major with increased PAH abundance, methods of tracing reddening that rely entirely on far-infrared emission (i.e., SFD, Planck-857, MF15) underestimate the correct extinction leading to an under density in the galaxy selection sample.
On the other hand, direct measurements of reddening using DESI and Gaia result in more uniform galaxy densities as function of PAH mass fraction, suggesting that these maps more accurately trace reddening. 
Similarly, WISE-W3, which traces PAH emission, has only a weak trend of galaxy under-density in areas of elevated PAH abundance.
{While the MAD value for the Gaia selected galaxy sample map (see Section~\ref{sec:galaxy_backlight_test}) implies more variations and inaccuracies in Gaia reddening map, the weak correlation of under-density with PAH abundance suggests that these variations are unlikely to be related to PAHs.}

For both WISE-W3 and Planck-857, we used the SFD dust temperature $T_{\rm SFD}$ to correct for varying dust emissivity. 
Any inaccuracies in this dust temperature will affect the recovered reddening.
Since denser regions correlate with higher PAH fraction, we note than an incorrect dust temperature correction can induce some artificial correlation between $q_{\rm PAH}$ and underestimates of extinction. 
This would subsequently produce under densities in the galaxy selection sample.
We discuss this effect in more detail in Section~\ref{sec:discussion}.

\section{Theoretical Modeling of PAH Mass Growth}
\label{sec:astrodust_pah_model}

Given the correlations between $q_{\rm PAH}$ and under-estimates of extinction by far-infrared emission based reddening maps, we consider theoretically the effect of changing PAH abundance and the ability of far-infrared emission to trace that change.
In particular, we consider the increase in PAH abundance due to the growth of PAHs via accretion.
Observational evidence for this growth process occurring toward Ursa Major has been seen in the form of correlations between lower $R_V$ values and higher $q_{\rm PAH}$~\citep{Zhang:2025a}.

To model the effect of PAH growth on reddening as well as on various indirect tracers of dust extinction, we adopt the astrodust+PAH dust model from \citet{Hensley:2023}.
\citet{Hensley:2023} constructed a unified model of interstellar dust components by fitting the dust model to observational constraints on extinction and emission at wavelengths ranging from $\sim0.1$\,$\mu$m~to~$\sim6800$\,$\mu$m .
This dust model consists of two components: an `astrodust' component and a PAH component.
The astrodust component comprises dust grains with a mixed composition that includes amorphous silicates and hydrocarbons \citep{Draine:2021d}.
The PAH component accounts for the vast majority of the smaller dust grains responsible for the mid-infrared emission \citep{Draine:2021b}.

To model the growth of PAHs, we begin with the best-fit grain size distribution for each component as the initial distributions.
While keeping the astrodust size distribution fixed, we then model the growth of PAHs via accretion of elements from the gas phase. 

\begin{figure*}[!ht]
    \centering
    \includegraphics[width=1\textwidth]{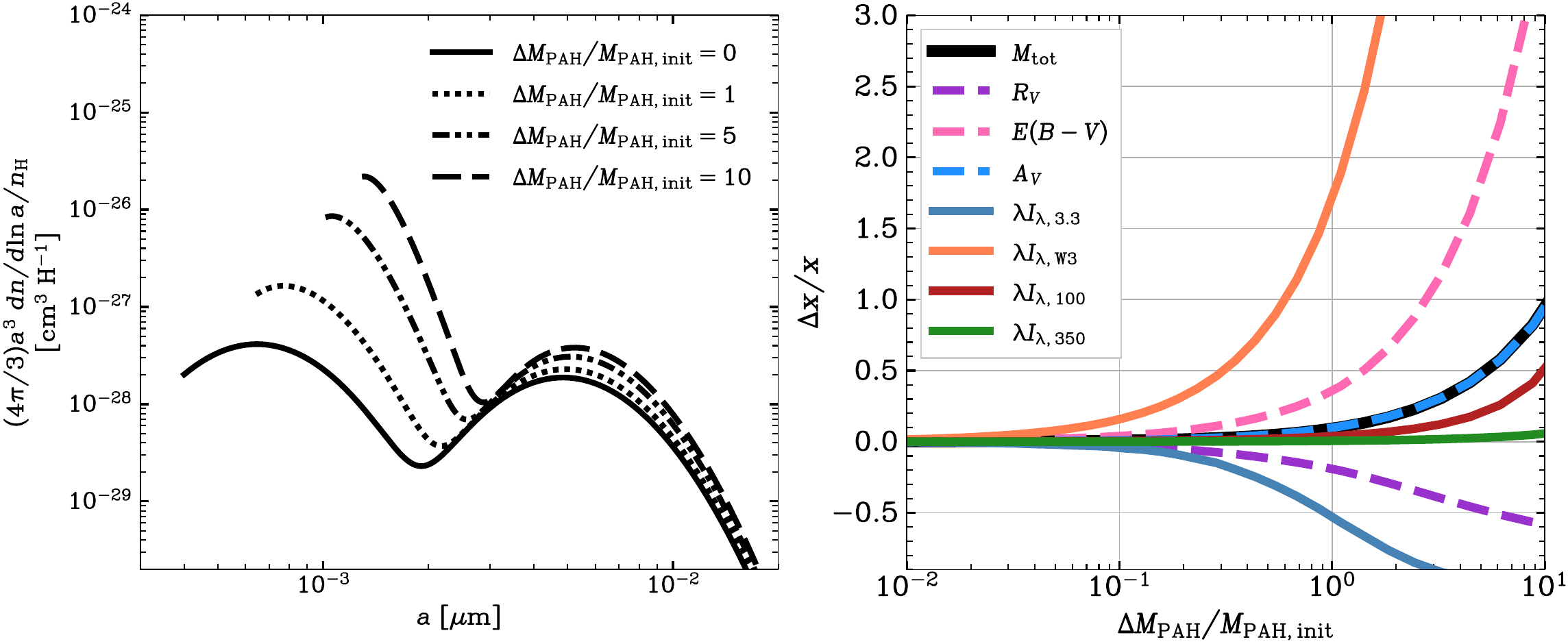}
    \caption{
    \textit{Left:} PAH dust grain size from the astrodust+PAH dust grain model \citep{Hensley:2023} as a function of effective radius $a$. 
    The solid black line indicates the initial PAH size distribution ($\Delta M_{\rm PAH} / M_{\rm PAH, init}=0$).
    Distributions with increased PAH mass due to accretion with $\Delta M_{\rm PAH} / M_{\rm PAH, init}=1, 5$, and 10 are plotted as  dotted, dashed-dotted, and dashed black lines, respectively.
    \textit{Right:} Change in extinction and emission as a function of increasing PAH mass ($\Delta M_{\rm PAH} / M_{\rm PAH, init}$). 
    The change in various extinction properties are shown as the dashed lines while the emission at various wavelengths are shown as the solid lines.
    As PAH mass grows, the emission at 100\,$\mu$m~and at 350\,$\mu$m~(857~GHz) are only weakly sensitive to the resulting increase in $E(B-V)$ while the emission 3.3\,$\mu$m~and in the WISE W3 band at 12 \,$\mu$m~are highly responsive.
    The process of PAH growth by accretion results in the loss of the relatively small PAH population that is responsible for the 3.3\,$\mu$m~emission. 
    This results in a decrease in emission with increasing PAH mass fraction.
    The WISE W3 band emission at 12 \,$\mu$m~largely originates from the comparatively larger PAHs.
    Unlike the emission at 3.3\,$\mu$m , emission in this band increases with increasing PAH mass fraction as accretion progresses.
    }
    \label{fig:accretion_model}
\end{figure*}

Our method of growing the PAHs via accretion follows that described in \citet{Zhang:2025a}.
In the simplest model of accretion, the effective radius $a$ of a grain grows over time $\Delta t$ by an amount $\Delta a$ that is independent of $a$ \citep{Hirashita:2012}.
The number of dust grains of a given effective radius $a$ is then given by
\begin{equation}
     n(a, t) = n(a - A(t), 0)
     ~,
\end{equation}
where $n(a,t)$ is the fraction of grains of with effective radius $a$ at time $t$ and $\Delta a = A(t)$ is the increase of grain size as a function of time.
Figure~\ref{fig:accretion_model}~(left) demonstrates the evolution of the PAH grain size distribution through accretion.
We denote the change in PAH mass $\Delta M_{\rm PAH}$.
The dashed line in Figure~\ref{fig:accretion_model}~(left), for example, represents the PAH size distribution after accreting from the gas phase ten times the initial PAH mass ($\Delta M_{\rm PAH} / M_{\rm PAH, init}=10$). 
Notably, as the dust grains grow via accretion, the smaller grains grow the most rapidly and the smallest PAHs are depleted without replenishment.

Figure~\ref{fig:accretion_model} (right) shows how emission and extinction evolve as the PAHs grow via accretion from the gas and the total PAH mass increases.
Naturally, we find that the growth of PAHs increases the extinction (as traced by the extinction in the $V$-band or $A_V$).
As previously shown in \citet{Zhang:2025a}, the growth of PAHs also reduces the value of $R_V$.
The strength of the prominent 2175 \AA~feature---which is thought to originate from PAHs or related carbonaceous nanoparticles---is proportional to the mass of PAHs. As a result, an increase in PAH mass, and thus the increase in the strength of the 2175 \AA~feature, sufficiently alters the slope of the extinction curve in optical/near-infrared (i.e., $R_V$).
Together, as $R_V \equiv A_V/E(B-V)$, these effects---the increase in $A_V$ and the decrease in $R_V$---results in an increase of the reddening $E(B-V)$. 
Growth of PAHs such that the total PAH mass in the dust doubles ($\Delta M_{\rm PAH}/M_{\rm PAH, init}= 1$), results in an approximate $20\%$ reduction in the value of $R_V$ (as was similarly modeled in \citealt{Zhang:2025a}) and a $35\%$ increase in $E(B-V)$ from the initial values.

In Figure~\ref{fig:accretion_model} (right), we show the evolution of emission at 3.3\,$\mu$m~and the emission integrated over the WISE~W3 band.
The near-infrared and mid-infrared wavelengths are home to prominent PAH emission features that trace the abundance of PAHs \citep[][Figure~3]{Hensley:2023}.
As a result, we expect both emission at 3.3\,$\mu$m~and the WISE~W3 band emission to be sensitive to the changes in PAH mass.
The WISE~W3 band emission clearly increases as the PAH mass increases.
While the 3.3\,$\mu$m~feature is sensitive to the PAH mass growth, the loss of small PAH grains in the accretion model leads to decrease in the strength of the 3.3\,$\mu$m~feature as PAH mass increases.
This is because it is the smallest PAHs that are able to attain temperatures hot enough to emit at 3.3\,$\mu$m~\citep{Draine:2001, Draine:2007}.
As accretion progresses, these smallest PAHs become less abundant even as the overall mass in PAHs increases.

For our modeling, we evolved the specific PAH size distribution from \citet{Hensley:2023} and followed the accretion description from \citet{Hirashita:2012}. Nonetheless, in general, we expect accretion to grow the sizes of the PAHs. 
PAHs can be produced by condensation in stellar outflows or built up from the gas phase in dense environments~\citep[e.g.,][]{Latter:1991, Sandstrom:2010, Burkhardt:2021, Lau:2022}.
As neither process is applicable in the relatively diffuse Ursa Major cloud, there is no mechanism to form new, smaller PAH molecules.
Without a method to replenish the smallest PAHs, we expect the decrease in 3.3\,$\mu$m~emission as accretion progresses to be a general consequence that is largely model-agnostic.

Finally, Figure~\ref{fig:accretion_model} (right) also shows how the far-infrared emission changes as PAH mass increases.
The far-infrared emission primarily originates in the larger dust grains that remain largely unchanged as PAH mass increases through accretion.
As a result, even significant increases in the PAH mass induce only a minor increase in the 100~$\mu$m or 857~GHz emission.
While we have fixed the astrodust component to its fiducial size distribution in this modeling, growth of these larger dust grains via these levels of accretion negligibly affects their size and thus extinction or emission properties~\citep[see][]{Zhang:2025a}.

\section{Joint Modeling of Extinction with Mid- and Far-Infrared Emission}
\label{sec:rv_modeling}
Since the far-infrared emission is largely insensitive to PAH abundance-induced extinction variations, a reddening map constructed with the addition of a tracer of PAH abundance should more accurately measure the true reddening than a far-infrared map alone.
To explore this in Ursa Major, we fit a simple parametric model of the contributions from mid-infrared emission and far-infrared emission to the observed reddening.
Given that the far-infrared emission is correlated with the non-PAH dust grains and the mid-infrared is correlated with the PAHs, each serves as proxy for their respective dust populations.

For the observed reddening, we use the Gaia measured $E(B-V)$ and $R_V$ values described in Section~\ref{sec:data} from \citet{Zhang:2025b}. 
We compute the observed extinction in the $V$-band $A_{V}^{\rm obs}$ as: 
\begin{equation}
    A_{V}^{\rm obs} = R_V^{\rm Gaia} \times E(B-V)^{\rm Gaia} {\rm ,}
\end{equation}
and the observed extinction in the $B$-band $A_{B}^{\rm obs}$ as:
\begin{equation}
    A_{B}^{\rm obs} = A_{V}^{\rm obs}  + E(B-V)^{\rm Gaia}\,.
\end{equation}

For the mid-infrared emission, we use the temperature-corrected mid-infrared WISE~W3 emission $I^{\rm W3,T}_\nu$. 
For the far-infrared emission, we use SFD reddening\footnote{This is equivalent to using the temperature-corrected IRAS 100\,$\mu$m~emission from \citet{Schlegel:1998} with a multiplicative factor.} $E(B-V)^{\rm SFD}$.

We model the contributions of each dust component as
\begin{align}
    A_{V}^{\rm} &= A_{V}^{\rm MIR}  I^{\rm W3,T}_\nu   + A_{V}^{\rm FIR} E(B-V)^{\rm SFD} \\
    A_{B}^{\rm} &= A_{B}^{\rm MIR}  I^{\rm W3,T}_\nu  +  A_{B}^{\rm FIR} E(B-V)^{\rm SFD}
\end{align}
where ${A_B^{\rm MIR}}$ and ${A_V^{\rm MIR}}$ denote the $A_B$ and $A_V$ contribution from the mid-infrared emission while ${A_B^{\rm FIR}}$ and ${A_V^{\rm FIR}}$ denote the $A_B$ and $A_V$ contribution from the far-infrared emission.

We use a Markov-Chain Monte Carlo (MCMC) to fit for the contributions to $A_B$ and $A_V$ from the mid-infrared emission and far-infrared emission. 
For this analysis, we employ the \texttt{emcee} package \citep{Foreman-Mackey:2013} which implements an affine invariant MCMC ensemble sampler. 
We perform the MCMC run with 32 walkers and 125,000 iterations with 20,000 iterations discarded as the burn-in.
We use broad priors for all four parameters enforcing only that the contributions to $A_B$ and $A_V$ for each component are positive.

\begin{figure*}[!hb]
\includegraphics[width=0.99\textwidth]{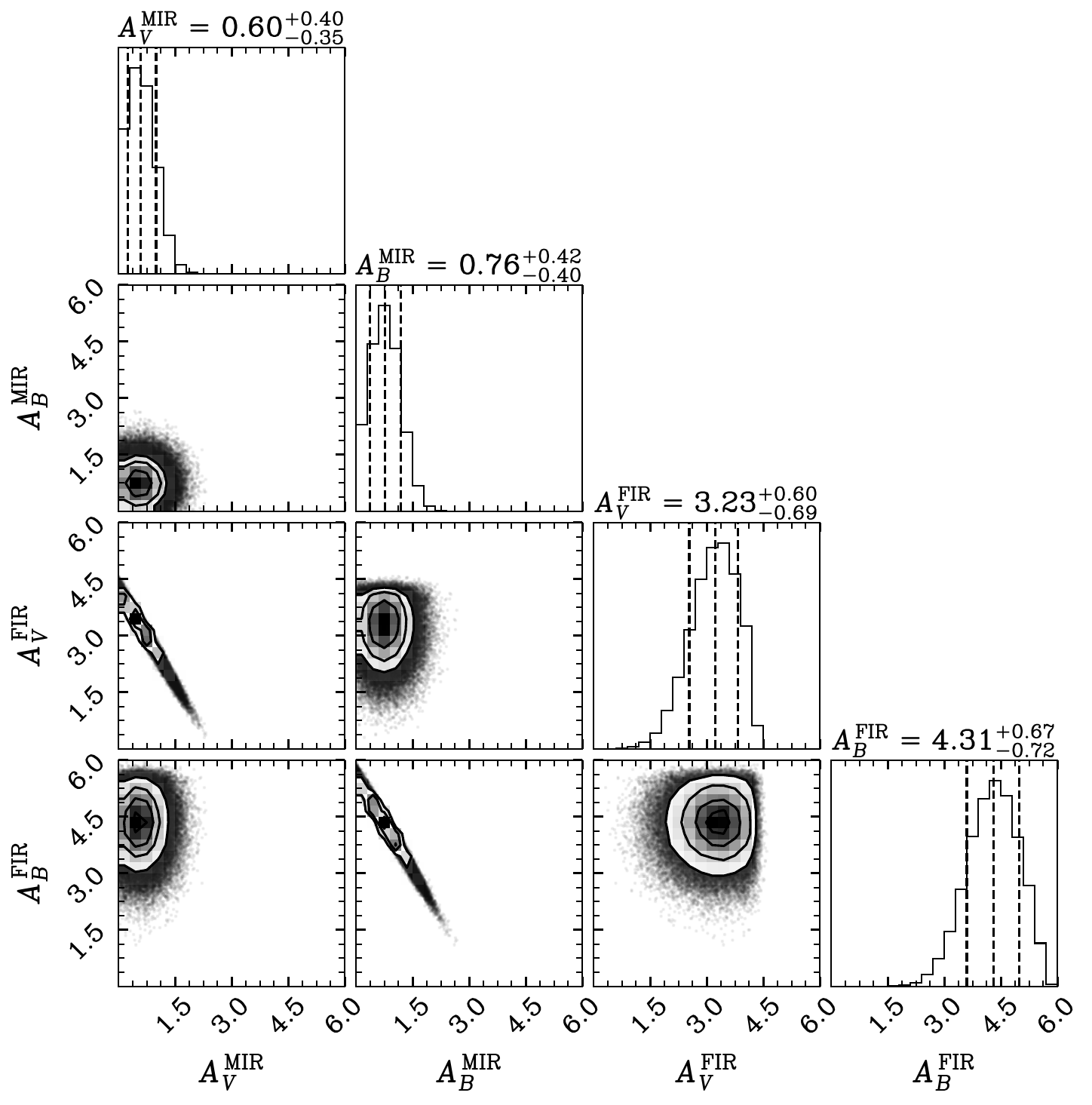}
\caption{Posterior distributions for the contributions from the mid-infrared and the far-infrared to $A_B$ and $A_V$.
$A_V^{\rm MIR}$ and $A_B^{\rm MIR}$ are in units of magnitudes per MJy\,sr$^{-1}$ while $A_V^{\rm FIR}$ and $A_B^{\rm FIR}$ are in units of magnitudes per $E(B-V)$ magnitude. 
The median of the marginal distributions and the upper and lower quantiles are shown.
\label{fig:avab_cornerplot}}
\end{figure*}

Figure~\ref{fig:avab_cornerplot} shows the posterior distributions for the contributions from the mid-infrared and the far-infrared to $A_B$ and $A_V$.
$A_V^{\rm MIR}$ and $A_B^{\rm MIR}$ are in units of magnitudes per MJy\,sr$^{-1}$ while $A_V^{\rm FIR}$ and $A_B^{\rm FIR}$ are in units of magnitudes per $E(B-V)$ magnitude.
Using the post-burn-in sampling, we compute the $R_V$ for each component (i.e, $R_V^{\rm MIR}$, $R_V^{\rm FIR}$).
{The median of the marginal posterior distribution for each component is $R_V^{\rm MIR}=0.29^{+2.17}_{-3.35}$ and $R_V^{\rm FIR}=2.25^{+4.10}_{-1.43}$.}
While these values are lower than the fiducial values from the fiducial astrodust+PAH model ($R_V^{\rm PAH}=0.71$, $R_V^{\rm AD}=4.40$), they remain consistent with PAH dust grains exhibiting a lower $R_V$.
Indeed, in our posterior sampling, $R_V^{\rm MIR} < R_V^{\rm FIR}$ for more than 75\% of our samples.

We note, however, the contribution of the mid-infrared emission ($A_V^{\rm MIR}$, $A_B^{\rm MIR}$) is only detected at a $\gtrsim1\sigma$ level (see Figure~\ref{fig:avab_cornerplot}).
Given that the 95\% credible interval includes zero, we cannot rule out the possibility that the extinction can be adequately explained without considering mid-infrared emission.
While the current evidence for Ursa Major alone does not strictly require inclusion of the mid-infrared component, a single component alone cannot produce the $R_V$ variations observed.
Furthermore, the low $R_V$ estimated for the mid-infrared data is aligned with model expectations \citep[e.g., ][]{Hensley:2023}. 
Given these considerations, there remains potential in deriving a more accurate reddening map from both mid-infrared and far-infrared emission.

\begin{figure}[!htp]
    \centering
    \includegraphics[width=0.99\columnwidth]{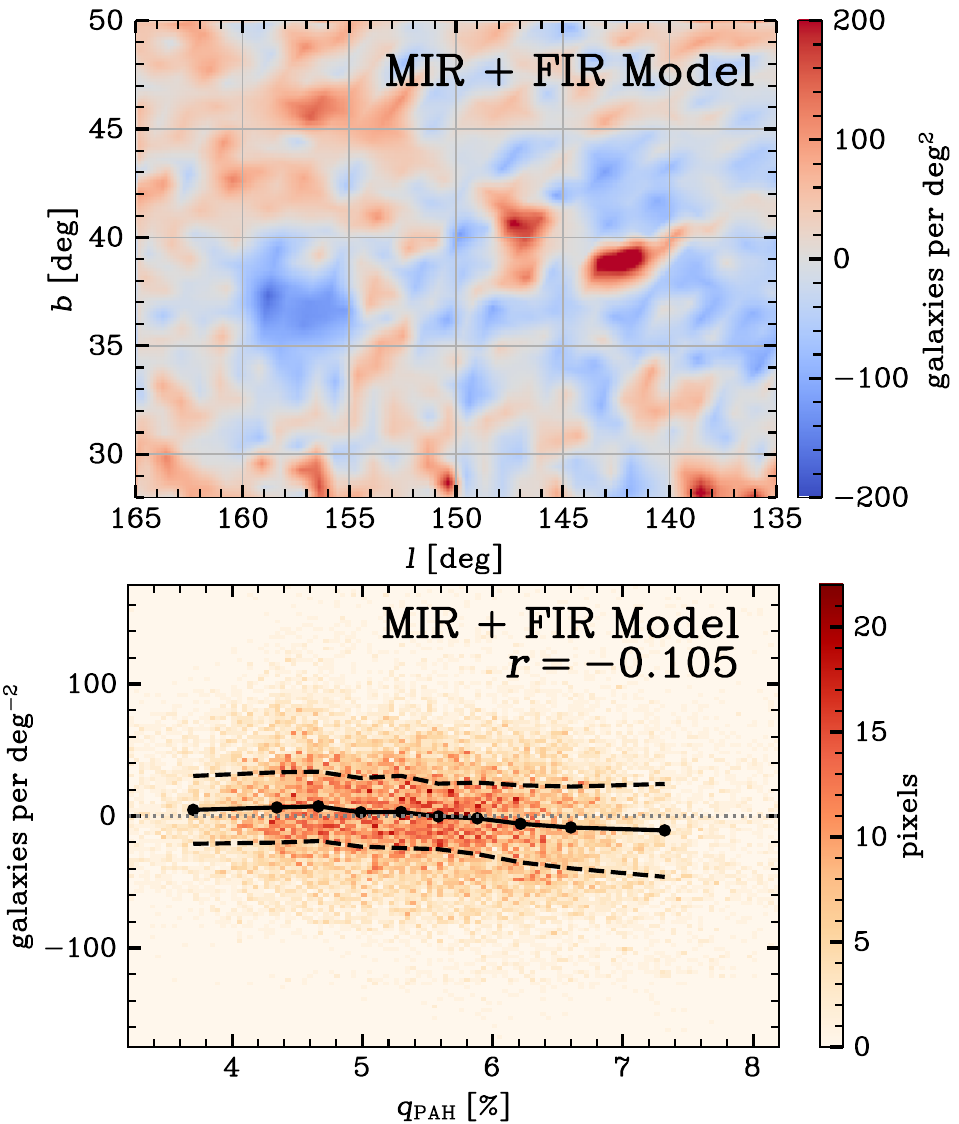}
    \caption{\textit{Top:} Galaxy density per deg$^2$ of the DESI ELG\_LOP galaxy sample selected using the extinction map constructed from both mid-infrared and far-infrared emission using the model from Section~\ref{sec:rv_modeling}.
    Compared with Figure~\ref{fig:galaxy_selection_map}, the analysis for this reddening maps qualitatively exhibits less spatial variation associated with Ursa Major cloud.
    \textit{Bottom:} Median-subtracted ELG\_LOP galaxy sample density vs.\,$q_{\rm PAH}$ for the model reddening map. The Pearson correlation coefficient $r$ is also shown.
    Compared with Figure~\ref{fig:galdense_v_qPAH}, this reddening map exhibits the weakest correlation of spatial variations with PAH abundance.
    }
    \label{fig:mirfir}
\end{figure}

To construct the reddening derived from both mid-infrared and far-infrared emission for the Ursa Major cloud, we use the maximum a posterior values of $A_V^{\rm MIR}$, $A_B^{\rm MIR}$, $A_V^{\rm FIR}$, and $A_B^{\rm FIR}$ to compute $E(B-V)$.
We then compare the performance of this new extinction map by using the galaxy density test described in Section~\ref{sec:galaxy_backlight_test}.
Figure~\ref{fig:mirfir} depicts the resulting galaxy sample map (top) as well as the correlations with the $q_{\rm PAH}$ (bottom).

Compared with the galaxy sample maps obtained using WISE-W3 and SFD individually, the galaxy density obtained using this extinction map qualitatively appears more spatially uniform.
Quantitatively, we also compute a lower MAD of 28 galaxies per deg$^2$ compared to 34 and 40 galaxies per deg$^2$ for WISE-W3 and SFD, respectively.
Furthermore, the jointly modeled extinction map shows the least correlation of the spatial galaxy density fluctuations with $q_{\rm PAH}$ (either positive or negative) of all the maps considered, suggesting that the deviations from uniformity are largely unrelated to PAHs.
Notably, we find that despite being fit to the Gaia map with a comparatively high MAD value, the joint mid-infrared and far-infrared model shows the smallest MAD value of all reddening maps.
Overall, the inclusion of a tracer of PAHs to jointly model the extinction appears to be an improvement over employing only far-infrared emission.

\section{Discussion}
\label{sec:discussion}

\subsection{PAHs as an Ultimate Limit for Far-infrared Extinction Maps}
\label{sec:pah_ult}

In the Ursa Major cloud complex, we consistently find that the far-infrared derived maps (i.e., SFD, Planck-857, MF15) underestimate the reddening compared to other methods in regions of elevated PAH abundance $q_{\rm PAH}$.
We employ the galaxy backlight test to validate the reddening maps, finding that reddening maps derived from far-infrared emission result in less uniform galaxy counts, suggesting that these maps are not accurately tracing the true reddening.
By contrast, the direct stellar reddening maps (i.e., DESI, Gaia) of Ursa Major both estimate higher values of reddening and produce galaxy sample map less correlated with Galactic structure, indicating a more accurate estimation of the reddening in the region.
The nonuniformities in galaxy counts observed for both DESI and Gaia exhibit weaker correlations with $q_{\rm PAH}$, implying that these reddening maps are not as susceptible to the same PAH-related limitations in tracing reddening seen in the far-infrared reddening maps.

Furthermore, the WISE-W3 reddening map does not suffer in the same manner as far-infrared maps.
Like the direct stellar reddening maps, the WISE-W3 reddening map appears to be more accurate---as demonstrated by the resulting galaxy samples uniformity in the galaxy backlight test---and has residuals that do not correlate as strongly with $q_{\rm PAH}$.
Given that WISE-W3 reddening is derived from observations that trace numerous PAH features in the mid-infrared, this is further evidence that the deficiencies of the far-infrared reddening maps is related to the abundance of PAHs.

While we argue that the DESI and Gaia reddening maps, by virtue of measuring reddening directly, are immune to errors induced by variations in PAH abundance, we do not imply that the stellar reddening maps are universally superior. 
For example, in our analysis, the DESI reddening map shows a positive trend with $q_{\rm PAH}$ that is the opposite to what is observed in the other reddening maps (see Figure~\ref{fig:galdense_v_qPAH}).
Similarly, for the Gaia reddening map, we find a relatively large variation in the ELG\_LOP galaxy density, as measured by the MAD value~(see Section~\ref{sec:galaxy_backlight_test}).
Given the limitations of stellar reddening maps as described in Section~\ref{sec:intro}, these may be related to increased systematics of such maps in the denser regions of the Ursa Major cloud.
Though understanding these limitations is valuable, these effects are likely unrelated to PAHs or any resulting PAH abundance induced extinction variations.

Prior work using measurements of $R_V$ has already shown that the Ursa Major is likely the site of PAH dust grain growth and increased PAH abundance \citep{Zhang:2025a}.
We find evidence that the $R_V$ of dust associated with PAH emission in Ursa Major is lower than for dust associated with emission from larger non-PAH dust grains.
This result is consistent with the theoretical predictions \citep[e.g.,][]{Weingartner:2001, Draine:2007, Guillet:2018}, providing further support that the PAH dust mass is indeed growing in Ursa Major.

The astrodust+PAH model predicts that changes in $R_V$ and $A_V$ are accompanied by an increase in $E(B-V)$ as PAHs grow by accretion of atoms from the gas.
However, the far-infrared emission primarily originates from the larger dust grains, and so increases in the PAH population leave the emission these longer wavelengths essentially unaffected.
Indeed, the PAH mass growth ($\Delta M_{\rm PAH} / M_{\rm PAH, init}\sim 1$) suggested by the $R_V$ variations ($\approx3.1\rightarrow2.5$) in Ursa Major causes a significant increase in $E(B-V)$ with only a modest change in emission at far-infrared wavelengths (see Figure~\ref{fig:accretion_model}).
Specifically, $\Delta M_{\rm PAH} / M_{\rm PAH, init}\sim 1$ results in a $35\%$ increase in $E(B-V)$, while only increasing emission at $100$\,$\mu$m~and $350$\,$\mu$m~by $\sim4\%$ and $0.6\%$, respectively. 
As a result, reddening maps constructed solely using far-infrared emission (e.g., SFD) will underestimate extinction in regions such as Ursa Major where there is an increased abundance of PAHs, exactly as is observed.

PAHs are not the only potential origin of reddening errors in far-infrared based maps.
For instance, the necessary dust temperature correction may be imperfect.
In principle, this temperature correction is designed to account for the that fact cooler dust emits less than an equivalent amount of warmer dust.
In our analysis, SFD and Planck-857 were corrected using a temperature map derived from DIRBE/COBE data and the MF15 was subjected to an effective temperature correction in their modeling.
If these temperature corrections are inaccurate, it could be insufficient to correct for the lower emissivity of the cooler dust in Ursa Major. 
In that case, a smaller quantity of dust would be erroneously estimated leading to an observed extinction underestimation.

However, a faulty temperature correction alone would not explain the variations seen in slope of the optical extinction curve $R_V$.
Increases in the PAH abundance naturally explains the $R_V$ variations as well as the underestimation of the reddening $E(B-V)$.
On the other hand, attributing all or most of the deficiencies in the SFD, Planck-857, and MF15 reddening maps to temperature variations would require a separate mechanism to explain the observed $R_V$ variations.

Nonetheless, this possibility highlights the fact that we have only focused on the effects from changes in the PAH abundance and distribution. 
PAHs represent the smallest of the dust grains in the Milky Way ISM---composing only a small fraction ($\sim$5\%) of the total interstellar dust by mass. 
It is possible, and indeed likely, that the properties of the larger grains, such as those modeled as `astrodust' in Section~\ref{sec:astrodust_pah_model}, vary spatially across the sky as well. 

In our study, we focused on the growth of PAHs through accretion as driving variations in PAH abundance (and thus variations in reddening).
However, it is likely that PAH abundance can vary via other mechanisms or for other reasons.
The destruction of PAHs is believed to occur in regions of strong ionizing radiation \citep{Montillaud:2013, Micelotta:2010}.
Observationally, this has been borne out by the lack of PAH emission in Galactic \ion{H}{2} \citep[e.g.,][]{Povich:2007}.
In a similar vein, large-scale studies of the cold neutral medium has suggested that PAHs prefer to exist in colder, relatively dense gas---again implying the destruction of PAH in more diffuse regions vulnerable to radiation~\citep{Hensley:2022a}.
More recently, studies of nearby galaxies have demonstrated a relationship between metallicity and PAH abundance, raising the possibility of such metallicity-related PAH variations in our own Milky Way \citep{Whitcomb:2024}.
As such, PAH growth via accretion need not be---and almost certainly is not---the sole cause for any PAH abundance-induced reddening changes.
PAH abundance variations for other reasons will likely still drive variations in reddening that commonly-used far-infrared based reddening maps are unable to detect.

\subsection{Future Observations with SPHEREx}
\label{sec:spherex} 
We explore the possibility of using future observations that trace the PAH abundance to augment existing far-infrared emission maps to construct extinction maps.
SPHEREx is a recently launched NASA observatory currently carrying out an all-sky spectral survey of the sky from 0.75\,$\mu$m~to 5\,$\mu$m~\citep{Crill:2020}. 
This all-sky survey will sensitively map the 3.3\,$\mu$m~PAH feature emission in the Milky Way.
As shown in Figure~\ref{fig:accretion_model}, theoretical modeling suggests that the 3.3\,$\mu$m~feature effectively traces increases in PAH mass fraction.

While we have used the WISE Band 3 emission to trace PAH abundance, measurements of the 3.3\,$\mu$m~PAH feature with SPHEREx is positioned to be more reliable.
In addition to having an angular resolution of $6''.2$ (comparable to WISE Band 3's angular resolution of $6''.5$), SPHEREx has a spectral resolving power of $35$ at 3.3\,$\mu$m. 
By contrast, Band 3 of WISE had a broad bandpass that---though it contains numerous PAH features---is more susceptible to contamination from line emission~\citep[e.g.,][]{Cluver:2017} or from the extragalactic background~\citep[e.g.,][]{Chiang:2019b}.
SPHEREx's spectral resolution will enable a more clean extraction and measurement of the 3.3\,$\mu$m~PAH feature and thus be a more precise tracer of PAH emission.

\begin{figure}
    \centering
    \includegraphics[width=0.99\columnwidth]{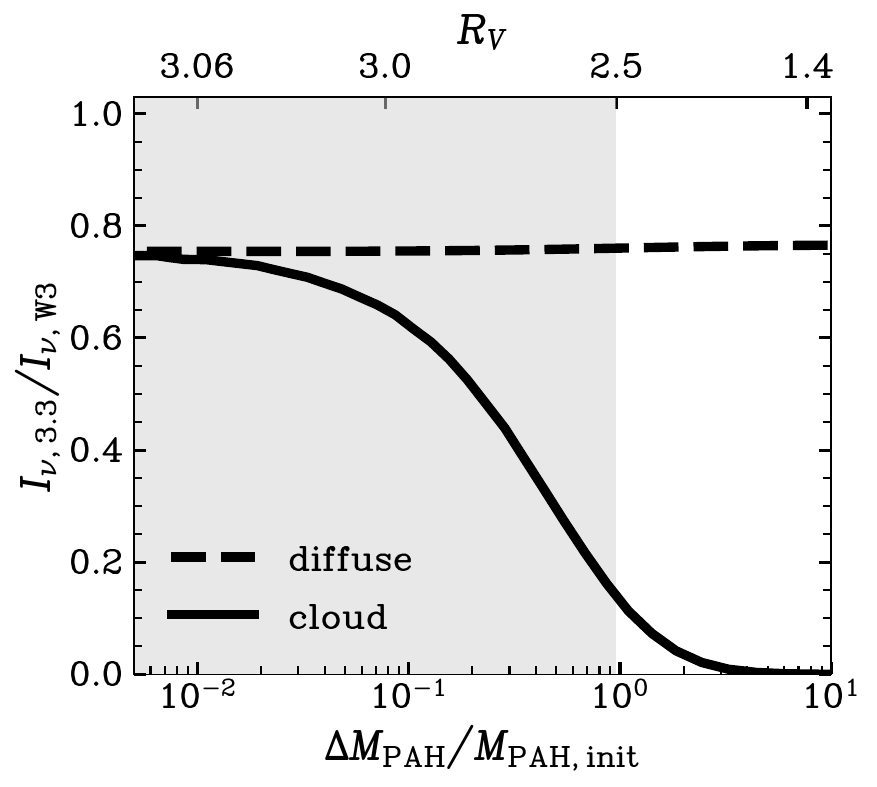}
    \caption{The ratio of 3.3\,$\mu$m~emission to emission at WISE~W3 band 
    $I_\nu^{3.3}/I_\nu^{\rm W3}$ as a function of the increase in PAH mass.
    The solid line indicates the accretion growth model shown in Figure~\ref{fig:accretion_model}.
    The dashed line indicates a model of PAH growth where small PAH grains are not lost.
    The grey band indicates the $R_V$ range where the Ursa Major region sees increased PAH grain growth.
    Depending on the behavior of the PAH dust grain distribution as the total PAH mass increases, the predicted $I_\nu^{3.3}$ may track $I_\nu^{\rm W3}$ closely or diverge due to the changes in grain size distribution.
    SPHEREx will be able to distinguish between these two possible models of PAH abundance variations.
    }
    \label{fig:33w3_evolution}
\end{figure}

Figure~\ref{fig:33w3_evolution} depicts the ratio of 3.3\,$\mu$m~emission to emission at WISE~W3 band $I_\nu^{3.3}/I_\nu^{\rm W3}$ as a function of the increase in PAH mass.
This value is sensitive both to the total mass fraction of PAHs as well as the precise PAH size distribution.
We depict two possible manners by which the PAH abundance can vary.

The first, indicated by the solid line in Figure~\ref{fig:33w3_evolution}, represents the scenario where PAH dust grains grow by accretion.
This is process described in Figure~\ref{fig:accretion_model}.
We designate this the ``cloud'' scenario as accretion growth requires condensation of elements from the gas that is likely to only occur in denser environments such as in clouds.

The second is indicated by the dashed line in Figure~\ref{fig:33w3_evolution}.
This represents a scenario where the total mass of the PAHs increases while maintaining the same size distribution---effectively by scaling the size distribution.
Such a scenario is likely more representative of PAH abundance variations in the diffuse ISM (as described in Section~\ref{sec:pah_ult}).
For example, a region near \ion{H}{2} region may have fewer PAHs than a region further away without possessing a substantially different PAH size distribution.
As a result, we designate this the ``diffuse'' scenario.

Due to this difference in the way the PAH size distributions change, the ratio $I_\nu^{3.3}/I_\nu^{\rm W3}$ evolves differently in each model. 
In the case of the diffuse model, the abundance of PAHs of all sizes increases at the same rate.
Thus, both the emission at 3.3\,$\mu$m~as well as WISE~W3 increases at a similar rate.
On the other hand, in the cloud model, the PAHs grow and smaller grains are not replenished.
The dust grains responsible for the 3.3\,$\mu$m~emission are the smaller, more neutral dust grains.
In the cloud model, these dust grains accrete and grow and, while the overall PAH mass fraction continues to increase, the fraction and quantity of those grains that emit strongly at 3.3\,$\mu$m~is decreased.
As a result, the loss of these grains results in $I_\nu^{3.3}$ becoming diminished compared to $I_\nu^{\rm W3}$.

Also marked in Figure~\ref{fig:33w3_evolution} is $R_V\approx2.5$, the lower limit of $R_V$ observed in Ursa Major, and the corresponding increase in PAH mass we infer.
While we expect that the cloud model will be more representative of the PAH evolution in Ursa Major, observations of the 3.3\,$\mu$m~PAH feature by SPHEREx will enable us to discern between these two different modes of PAH abundance variations through the ratio of the PAH features as $R_V$ decreases in this region.

Notably, measurement of the 3.3\,$\mu$m~PAH feature remains a promising tracer of PAH growth not traced by far-infrared emission irrespective of exactly how the PAH dust grain distribution evolves as the mass increases.
{Jointly modeling the extinction of Ursa Major with mid-infrared and far-infrared emission shows a marked improvement compared to relying solely on far-infrared emission or solely on mid-infrared emission (see Figure~\ref{fig:mirfir}).
By tracing both the PAH population and the larger non-PAH dust grain population, such an extinction map appears capable of accounting for extinction variations that may be driven by PAH abundance changes.}
Observations of the 3.3\,$\mu$m~PAH feature by SPHEREx is therefore a promising avenue to address the limitations of far-infrared observations in tracing dust extinction.

\section{Conclusion}
\label{sec:conclusion}
In this work, we demonstrate that reddening derived from far-infrared emission maps are insensitive to reddening variations induced by changes in PAH abundance.
Comparing reddening maps constructed using various methods, we find that far-infrared based reddening maps such as SFD underestimate the extinction in the Ursa Major cloud.
On the other hand, reddening maps derived from direct stellar reddening measurements and mid-infrared emission from WISE appear more robust to changes in PAH mass fraction. 
Jointly modeling the dust extinction with mid-infrared and far-infrared emission also results in a qualitatively and quantitatively improved map, when compared to using far-infrared emission alone.

Although we have focused on the growth of PAHs through accretion as the reason for PAH variations, changes in PAH abundance can result from other processes.
As such, while the exact conditions of PAH accretion in Ursa Major are not guaranteed to be prevalent throughout the Milky Way, PAH abundance variations due to other processes may still lead to variations in reddening that far-infrared based reddening maps are unable to detect. 

Accounting for changes in the PAH mass fraction requires a sensitive tracer of the PAH population.
Near and mid-infrared wavelengths are home to numerous PAH emission features including the 3.3\,$\mu$m~PAH feature.
We demonstrate that SPHEREx, which is currently mapping this feature across the entire sky, presents a promising method of addressing this limitation.

\begin{acknowledgments}
This research was carried out at the Jet Propulsion Laboratory, California Institute of Technology, under a contract with the National Aeronautics and Space Administration (80NM0018D0004). 
Some of the results in this paper have been derived using the healpy and HEALPix package.

The Legacy Surveys consist of three individual and complementary projects: the Dark Energy Camera Legacy Survey (DECaLS; Proposal ID \#2014B-0404; PIs: David Schlegel and Arjun Dey), the Beijing-Arizona Sky Survey (BASS; NOAO Prop. ID \#2015A-0801; PIs: Zhou Xu and Xiaohui Fan), and the Mayall z-band Legacy Survey (MzLS; Prop. ID \#2016A-0453; PI: Arjun Dey). DECaLS, BASS and MzLS together include data obtained, respectively, at the Blanco telescope, Cerro Tololo Inter-American Observatory, NSF’s NOIRLab; the Bok telescope, Steward Observatory, University of Arizona; and the Mayall telescope, Kitt Peak National Observatory, NOIRLab. Pipeline processing and analyses of the data were supported by NOIRLab and the Lawrence Berkeley National Laboratory (LBNL). The Legacy Surveys project is honored to be permitted to conduct astronomical research on Iolkam Du’ag (Kitt Peak), a mountain with particular significance to the Tohono O’odham Nation.

NOIRLab is operated by the Association of Universities for Research in Astronomy (AURA) under a cooperative agreement with the National Science Foundation. LBNL is managed by the Regents of the University of California under contract to the U.S. Department of Energy.

This project used data obtained with the Dark Energy Camera (DECam), which was constructed by the Dark Energy Survey (DES) collaboration. Funding for the DES Projects has been provided by the U.S. Department of Energy, the U.S. National Science Foundation, the Ministry of Science and Education of Spain, the Science and Technology Facilities Council of the United Kingdom, the Higher Education Funding Council for England, the National Center for Supercomputing Applications at the University of Illinois at Urbana-Champaign, the Kavli Institute of Cosmological Physics at the University of Chicago, Center for Cosmology and Astro-Particle Physics at the Ohio State University, the Mitchell Institute for Fundamental Physics and Astronomy at Texas A\&M University, Financiadora de Estudos e Projetos, Fundacao Carlos Chagas Filho de Amparo, Financiadora de Estudos e Projetos, Fundacao Carlos Chagas Filho de Amparo a Pesquisa do Estado do Rio de Janeiro, Conselho Nacional de Desenvolvimento Cientifico e Tecnologico and the Ministerio da Ciencia, Tecnologia e Inovacao, the Deutsche Forschungsgemeinschaft and the Collaborating Institutions in the Dark Energy Survey. The Collaborating Institutions are Argonne National Laboratory, the University of California at Santa Cruz, the University of Cambridge, Centro de Investigaciones Energeticas, Medioambientales y Tecnologicas-Madrid, the University of Chicago, University College London, the DES-Brazil Consortium, the University of Edinburgh, the Eidgenossische Technische Hochschule (ETH) Zurich, Fermi National Accelerator Laboratory, the University of Illinois at Urbana-Champaign, the Institut de Ciencies de l'Espai (IEEC/CSIC), the Institut de Fisica d’Altes Energies, Lawrence Berkeley National Laboratory, the Ludwig Maximilians Universitat Munchen and the associated Excellence Cluster Universe, the University of Michigan, NSF’s NOIRLab, the University of Nottingham, the Ohio State University, the University of Pennsylvania, the University of Portsmouth, SLAC National Accelerator Laboratory, Stanford University, the University of Sussex, and Texas A\&M University.

BASS is a key project of the Telescope Access Program (TAP), which has been funded by the National Astronomical Observatories of China, the Chinese Academy of Sciences (the Strategic Priority Research Program “The Emergence of Cosmological Structures” Grant \# XDB09000000), and the Special Fund for Astronomy from the Ministry of Finance. The BASS is also supported by the External Cooperation Program of Chinese Academy of Sciences (Grant \# 114A11KYSB20160057), and Chinese National Natural Science Foundation (Grant \# 12120101003, \# 11433005).

The Legacy Survey team makes use of data products from the Near-Earth Object Wide-field Infrared Survey Explorer (NEOWISE), which is a project of the Jet Propulsion Laboratory/California Institute of Technology. NEOWISE is funded by the National Aeronautics and Space Administration.

The Legacy Surveys imaging of the DESI footprint is supported by the Director, Office of Science, Office of High Energy Physics of the U.S. Department of Energy under Contract No. DE-AC02-05CH1123, by the National Energy Research Scientific Computing Center, a DOE Office of Science User Facility under the same contract; and by the U.S. National Science Foundation, Division of Astronomical Sciences under Contract No. AST-0950945 to NOAO.

\end{acknowledgments}

\software{
    astropy \citep{AstropyCollaboration:2013, AstropyCollaboration:2018, AstropyCollaboration:2022},
    numpy \citep{harris2020array}, 
    scipy \citep{Virtanen:2020}, 
    matplotlib \citep{Hunter:2007},
    healpy \citep{Zonca:2019},
    emcee \citep{Foreman-Mackey:2013},
    corner.py \citep{Foreman-Mackey:2016}
}
\bibliography{references, software, Planck_bib}{}
\bibliographystyle{aasjournalv7}

\end{document}